\documentclass[11pt]{article}

\usepackage[margin=1in]{geometry}
\usepackage{graphicx}

\usepackage{alltt}

\usepackage{amsmath}

\usepackage{upgreek}
\usepackage{amssymb}
\usepackage{enumitem}

\usepackage{xcolor} 
\usepackage{url}
\usepackage[export]{adjustbox}
\usepackage[singlelinecheck=false]{caption}

\DeclareMathOperator{\E}{E}
\DeclareMathOperator{\Var}{var}
\DeclareMathOperator{\PPp}{{P}_{p}}
\DeclareMathOperator{\Ep}{{E}_{p}}
\DeclareMathOperator{\Varp}{{var}_{p}}
\DeclareMathOperator{\MSEp}{{MSE}_{p}}
\DeclareMathOperator{\mse}{mse}
\DeclareMathOperator{\mseu}{mse_u}
\DeclareMathOperator{\mseb}{mse_b}
\DeclareMathOperator{\RMSE}{RMSE}
\DeclareMathOperator{\AB}{AB}
\DeclareMathOperator*{\argmin}{arg\,min}

\def\leq{\leqslant }
\def\geq{\geqslant }

\def\boldhline{\noalign{\global\arrayrulewidth.8pt}\hline\noalign{\global\arrayrulewidth.4pt}}

\newcommand{\pkg}[1]{\texttt{#1}}

\begin{document}

\LARGE
\begin{center}
\textbf{Design-based composite estimation of small proportions in small domains}\\[1.75\baselineskip]

\normalsize

\text{Andrius {\v C}iginas}\\[1.75\baselineskip]

{\footnotesize
Vilnius University \\
[2.25\baselineskip]}
\end{center}

\begin{abstract}

\normalsize

Traditional direct estimation methods are not efficient for domains of a survey population with small sample sizes. To estimate the domain proportions, we combine the direct estimators and the regression-synthetic estimators based on domain-level auxiliary information. For the case of small true proportions, we introduce the design-based linear combination that is a robust alternative to the empirical best linear unbiased predictor (EBLUP) based on the Fay--Herriot model. We also consider an adaptive procedure optimizing a sample-size-dependent composite estimator, which depends on a single parameter for all domains. 

We imitate the Lithuanian Labor Force Survey, where we estimate the proportions of the unemployed and employed in municipalities. We show where the considered design-based compositions and estimators of their mean square errors are competitive for EBLUP and its accuracy estimation. \vskip 2mm 

\textbf{Keywords:} small area estimation, area-level model, composite estimator, sample-size-dependent estimator, Labor Force Survey.

\end{abstract}

\normalsize

\section{Introduction}\label{s:1} 

Design-based and model-assisted direct estimators of parameters rely only on the sample of the estimation domain (area). Therefore, after the sample is selected, their application for some unplanned domains leads to high variances of the estimators because of too small sample sizes. In the small area estimation theory \cite{RM_2015}, indirect estimators borrow sample information from neighbor domains through auxiliary information and linking models. These model-based estimators usually have lower variances than the direct estimators, but their biases can be relatively large.

To estimate proportions in the domains, one can consider explicit linking models based on auxiliary data aggregated to the domain level. A popular model is the Fay--Herriot (FH) model, which is a separate case of linear mixed models, and the empirical best linear unbiased predictors (EBLUPs) of the domain means or proportions are derived from it \cite{FH_1979}. That small area predictor is expressed as the linear combination of a regression-synthetic estimator and the direct estimator. While the former part accounts for a variation reflected in the auxiliary data, the direct component exploits the unbiasedness property. Compositions of the synthetic and the direct estimators constitute an important class of indirect estimators. Before the mixed models, traditional design-based composite estimators were often used \cite[Chapter 3]{RM_2015}. However, now it is accepted that the models including random area-specific effects are more useful. For example, they are more convenient to handle complex data structures than the traditional estimators with only randomness induced by the sampling design. Another notable drawback of the latter estimators is the difficulty in estimating their precision. The problem is with bias estimation, while it is well elaborated for the estimators like EBLUP. 

We use a conditional analysis to construct the design-based composite estimator, which is in some sense similar to EBLUP. According to the construction, it is a robust estimator suitable for small or large domain proportions. For the comparison, we also consider the sample-size-dependent (SSD) compositions introduced in \cite{DSC_1982}. To optimize the latter estimators with respect to their parameter, we apply the strategy based on the minimization of estimated average mean square error (MSE) as proposed in \cite{C_2020}. The MSEs of both the design-based compositions are estimated as suggested in \cite{C_2021}.

We compare the estimators and their MSE estimators in the simulation study using the Lithuanian Labor Force Survey (LFS) data, where fractions of the unemployed and employed are the proportions of interest estimated in municipalities. Applications of EBLUPs to LFS unemployment data of other countries are found, for example, in \cite{BBBKS_2008, GLMMS_2010, MS_2019}. SSD compositions, with subjectively chosen values of the parameter, are used in \cite{DSC_1982, UGMS_2009}. The adaptive selection of values of this parameter is applied to estimate the proportions of unemployed in \cite{C_2020}.

\section{Basic assumptions and direct estimation}\label{s:2}

The set ${\cal U}=\{1, \ldots, N \}$ consists of the labels of elements of the survey population. Let $y$ be a binary study variable with the fixed values $y_1,\ldots, y_N$ assigned to the corresponding elements. To estimate the proportions in the population and its subsets, the sample $s\subset {\cal U}$ of size $n<N$ is drawn by the sampling design $p(\cdot)$, and $\pi_k=\PPp\{ k\in s\}>0$, $ k\in {\cal U}$, are inclusion into the sample probabilities. Here the symbol $\PPp$, and hereafter $\Ep$, $\Varp$, and $\MSEp$ denote probability, expectation, variance, and MSE according to $p(\cdot)$, respectively. The characteristic $\Varp (\cdot)$ is called the sampling variance or design variance. 

Let ${\cal U}={\cal U}_1\cup\cdots\cup{\cal U}_M$ be the partition of the population into the non-overlapping domains, where the domain ${\cal U}_i$ contains $N_i$ elements. Then the domain sample $s_i=s\cap {\cal U}_i$ is of size $n_i\leq N_i$. We aim to estimate the proportions
\begin{equation}\label{target_means}
\theta_i=\frac{1}{N_i}\sum_{k\in {\cal U}_i}y_k, \qquad i=1,\ldots, M,
\end{equation}
where the numbers $N_i$ are assumed to be known. If the design $p(\cdot)$ does not ensure the fixed sizes $n_i$, then they can be too small to get suficiently accurate direct estimates $\hat\theta_i^{\mathrm d}$ of \eqref{target_means}.

Assume that, for each domain ${\cal U}_i$, the auxiliary information is available as the vector of known characteristics $\mathbf{z}_i=(z_{i1}, z_{i2},\ldots, z_{iP})'$. This assumption narrows a choice of direct estimators to the design unbiased Horvitz--Thompson estimators $\hat \theta_i^{\mathrm {HT}}=N_i^{-1}\sum_{k\in s_i}y_k/\pi_k$ of $\theta_i$ or the weighted sample proportions
\begin{equation}\label{dir_ratio}
\hat \theta_i^{\mathrm H}=\frac{1}{\widehat N_i}\sum_{k\in s_i}\frac{y_k}{\pi_k}, \quad \text{where} \quad \widehat N_i=\sum_{k\in s_i}\frac{1}{\pi_k}, \qquad i=1,\ldots, M,
\end{equation}
that are approximately unbiased. The approximate sampling variances of \eqref{dir_ratio} and their estimators have the expressions \cite[p. 185]{SSW_1992}
\begin{equation}\label{dir_ratio_var}
\Varp(\hat \theta_i^{\mathrm H}) \approx \psi_i^{\mathrm H}=\frac{1}{N_i^2} \sum_{k\in {\cal U}_i} \sum_{l\in {\cal U}_i} (\pi_{kl}-\pi_k\pi_l) \frac{(y_k-\theta_i)(y_l-\theta_i)}{\pi_k\pi_l}, \qquad i=1,\ldots, M,
\end{equation} 
and 
\begin{equation}\label{dir_ratio_var_est}
\hat \psi_i^{\mathrm H}=\frac{1}{\widehat N_i^2} \sum_{k\in s_i} \sum_{l\in s_i} (1-\pi_k\pi_l/\pi_{kl}) \frac{(y_k-\hat \theta_i^{\mathrm H})(y_l-\hat \theta_i^{\mathrm H})}{\pi_k\pi_l}, \qquad i=1,\ldots, M,
\end{equation}
respectively, where $\pi_{kl}=\PPp\{ k, l\in s\}>0$ is the probability that both of the elements $k$ and $l$ will be included into the sample.

\section{EBLUP under the Fay--Herriot model}\label{s:3}

The direct estimators $\hat\theta_i^{\mathrm d}$ of the domain proportions can be improved using the FH model \cite{FH_1979}. The data for this domain-level model are the estimates $\hat\theta_i^{\mathrm d}$, their corresponding estimates $\hat \psi_i$ of the sampling variances $\psi_i=\Varp(\hat \theta_i^{\mathrm d})$, and the covariates $\mathbf{z}_i$, $ i=1,\ldots, M$. The basic FH model consists of two parts, see \cite[Section 4.2]{RM_2015}, that are combined into the linear mixed model
\begin{equation}\label{M1_eblup}
\hat\theta_i^{\mathrm d}=\mathbf{z}_i' \boldsymbol{\upbeta}+v_i+\varepsilon_i, \qquad i=1,\ldots, M,
\end{equation}
where $\boldsymbol{\upbeta}=(\beta_1,\ldots, \beta_P)'$ is the vector of fixed effects, the sampling errors $\varepsilon_i$ are assumed independent with $\Ep (\varepsilon_i)=0$ and $\Varp (\varepsilon_i)=\psi_i$, and random domain effects $v_i$ are assumed independent of these errors. The latter effects are supposed to be independent and identically distributed with $\E (v_i)=0$ and $\Var (v_i)=\sigma_v^2\geq 0$ in respect of a distribution, different from that generated by the design $p(\cdot)$.

Treating the estimates $\hat \psi_i$ as given numbers, the method of EBLUP leads to the predictions of domain proportions \eqref{target_means} that are expressed as the linear combinations \cite{FH_1979}
\begin{equation}\label{eblup}
\hat\theta_i^{\mathrm {FH}}=\hat\theta_i^{\mathrm {FH}}(\hat \psi_i)=\hat\gamma_i\hat\theta_i^{\mathrm d}+(1-\hat\gamma_i)\mathbf{z}_i' \boldsymbol{\hat\upbeta}, \quad \text{with} \quad \hat\gamma_i=\frac{\hat\sigma_v^2}{\hat \psi_i+\hat\sigma_v^2}, \qquad i=1,\ldots, M,
\end{equation}
and
\begin{equation*}
\boldsymbol{\hat\upbeta}=\Biggl( \sum_{i=1}^M \frac{\mathbf{z}_i\mathbf{z}_i'}{\hat \psi_i+\hat\sigma_v^2}\Biggr)^{-1} \sum_{i=1}^M \frac{\mathbf{z}_i \hat\theta_i^{\mathrm d}}{\hat \psi_i+\hat\sigma_v^2},
\end{equation*}
where $\hat\sigma_v^2$ is an estimator of the variance $\sigma_v^2$. One of the ways to estimate $\sigma_v^2$ is the estimator $\hat\sigma_v^2$ based on the method of moments, as originally proposed by \cite{FH_1979}. For this estimator, approximately unbiased estimators of MSEs of \eqref{eblup} were derived in \cite{DRS_2005}:
\begin{equation}\label{mse_eblup}
\begin{split}
\mse (\hat\theta_i^{\mathrm {FH}})= \hat\gamma_i\hat \psi_i + (1-\hat\gamma_i)^2  & \Biggl[ \mathbf{z}_i' \Biggl( \sum_{j=1}^M \frac{\mathbf{z}_j\mathbf{z}_j'}{\hat \psi_j+\hat\sigma_v^2}\Biggr)^{-1} \mathbf{z}_i + \frac{4M}{\hat \psi_i+\hat\sigma_v^2} \Biggl( \sum_{j=1}^M \frac{1}{\hat \psi_j+\hat\sigma_v^2}\Biggr)^{-2} \Biggr. \\
& \Biggl. -2\hat\sigma_v^2 \Biggl( \sum_{j=1}^M \hat\gamma_j \Biggr)^{-3} \Biggl\{ M\sum_{j=1}^M \hat\gamma_j^2 - \Biggl( \sum_{j=1}^M \hat\gamma_j \Biggr)^2 \Biggr\} \Biggr] , \qquad i=1,\ldots, M.
\end{split}
\end{equation} 

Predictions \eqref{eblup} and their MSE estimators \eqref{mse_eblup} depend also on the estimators $\hat \psi_i$ of the sampling variances $\psi_i$ of $\hat\theta_i^{\mathrm d}$. However, direct estimators $\hat \psi_i^{\mathrm d}$ of $\psi_i$, as, for example, approximately design unbiased estimators \eqref{dir_ratio_var_est} of \eqref{dir_ratio_var} for $\hat \theta_i^{\mathrm H}$, have large variances themselves for small sample sizes. Therefore, the direct estimates $\hat \psi_i^{\mathrm d}$ are smoothed and new more stable estimates $\hat \psi_i^{\mathrm s}$ are used in \eqref{eblup} and \eqref{mse_eblup}. It is called the generalized variance function (GVF) approach \cite{W_2007}. The specific example of the GVF method, similar to that used for estimation of census undercounts in \cite{D_1995}, is to assume that $\psi_i \approx KN_i^\gamma$ and estimate the parameters $K>0$ and $\gamma\in \mathbb R$ using the regression model
\begin{equation*}
\log(\hat \psi_i^{\mathrm d})=\log(K)+\gamma\log(N_i)+\eta_i, \qquad i=1,\ldots, M,
\end{equation*}
where errors $\eta_i$ are independent and identically distributed. That is, the smoothed estimates 
\begin{equation}\label{smooth_Dick}
\hat \psi_i^{\mathrm {sD}}=\widehat K N_i^{\hat\gamma}, \qquad i=1,\ldots, M,
\end{equation} 
of $\psi_i$ are based on the ordinary least squares estimates of the regression parameters. Other smoothing examples are pooled variance estimation \cite{BBBKS_2008} and a nonparametric smoothing like in \cite{GLMMS_2010}. Despite the smoothing, estimators \eqref{mse_eblup} tend to underestimate MSEs of \eqref{eblup} because the estimation of the sampling variances $\psi_i$ is ignored in the derivation of \eqref{mse_eblup}.

\section{Design-based composite estimation}\label{s:4}

\subsection{Evaluation of optimal compositions and their accuracy estimation}\label{s:41} 

Let us exclude the random effects $v_i$ from FH model \eqref{M1_eblup}. Then the model becomes
\begin{equation}\label{M1}
\hat\theta_i^{\mathrm d}=\mathbf{z}_i' \boldsymbol{\upbeta}+\varepsilon_i, \qquad i=1,\ldots, M,
\end{equation}
and, using the estimates $\hat \psi_i$ of the variances $\psi_i$, we arrive to the regression-synthetic estimators
\begin{equation}\label{Rsyn}
\hat\theta_i^{\mathrm S}=\hat\theta_i^{\mathrm S}(\hat \psi_i)=\mathbf{z}_i' \boldsymbol{\hat\upbeta}, \qquad i=1,\ldots, M,
\end{equation}
of the domain proportions $\theta_i$, where 
\begin{equation}\label{beta_est}
\boldsymbol{\hat\upbeta}=\Biggl( \sum_{i=1}^M \frac{\mathbf{z}_i\mathbf{z}_i'}{\hat \psi_i}\Biggr)^{-1} \sum_{i=1}^M \frac{\mathbf{z}_i \hat\theta_i^{\mathrm d}}{\hat \psi_i}
\end{equation}
is the generalized least squares estimate of $\boldsymbol{\upbeta}$. Here, as for EBLUPs, the use of smoothed estimates $\hat \psi_i=\hat \psi_i^{\mathrm s}$ instead of $\hat \psi_i^{\mathrm d}$ stabilizes synthetic estimators \eqref{Rsyn}. 

Estimators \eqref{Rsyn} rely on a synthetic assumption that the parameter $\boldsymbol{\upbeta}$ is the same across all domains. Therefore, having a good regression model, their sampling variances are small, compared to that of the direct estimators $\hat\theta_i^{\mathrm d}$ or even the EBLUPs $\hat\theta_i^{\mathrm {FH}}$. However, the design biases of \eqref{Rsyn} can be relatively large if the synthetic assumption is not realistic. To find a trade-off between larger variances of $\hat\theta_i^{\mathrm d}$ and biases of the synthetic estimators $\hat\theta_i^{\mathrm S}$, we consider their linear combinations
\begin{equation}\label{gen_lin_comb}
\tilde\theta_i^{\mathrm C}=\tilde\theta_i^{\mathrm C}(\lambda_i)=\lambda_i\hat\theta_i^{\mathrm d}+(1-\lambda_i)\hat\theta_i^{\mathrm S}, \qquad i=1,\ldots, M,
\end{equation}
with weights $0\leq \lambda_i\leq 1$. Minimizing the function $\MSEp(\tilde\theta_i^{\mathrm C}(\lambda_i))$ with respect to $\lambda_i$, the optimal weight for the domain ${\cal U}_i$ is the population parameter \cite[Section 3.3]{RM_2015}
\begin{equation}\label{opt_lambda}
\lambda_i^*=\frac{\MSEp(\hat\theta_i^{\mathrm S})-C_i}{\MSEp(\hat\theta_i^{\mathrm d})+\MSEp (\hat\theta_i^{\mathrm S})-2C_i} \quad \text{with} \quad C_i=\Ep(\hat\theta_i^{\mathrm d}-\theta_i)(\hat\theta_i^{\mathrm S}-\theta_i).
\end{equation} 
Assuming that $|C_i| \ll \MSEp(\hat\theta_i^{\mathrm S})$, the approximation $\lambda_i^*\approx \MSEp(\hat\theta_i^{\mathrm S})/(\MSEp(\hat\theta_i^{\mathrm d})+\MSEp (\hat\theta_i^{\mathrm S}))$ is applied, but the further difficulty is to evaluate the quantities $\MSEp(\hat\theta_i^{\mathrm S})$. A common approach to this is to use the representation \cite[Section 3.2.5]{RM_2015}
\begin{equation}\label{GW}
\MSEp(\hat\theta_i^{\mathrm S})=\Ep (\hat\theta_i^{\mathrm S}-\hat\theta_i^{\mathrm d})^2-\Varp(\hat\theta_i^{\mathrm S}-\hat\theta_i^{\mathrm d})+\Varp(\hat\theta_i^{\mathrm S}),
\end{equation}
where $\hat\theta_i^{\mathrm d}$ is assumed to be unbiased, and then to build an approximately design unbiased estimator
\begin{equation}\label{GWest}
\mseu (\hat\theta_i^{\mathrm S})=(\hat\theta_i^{\mathrm S}-\hat\theta_i^{\mathrm d})^2-\hat\sigma^2(\hat\theta_i^{\mathrm S}-\hat\theta_i^{\mathrm d})+\hat\sigma^2(\hat\theta_i^{\mathrm S})
\end{equation} 
of \eqref{GW}, where $\hat\sigma^2(\cdot)$ is an estimator of the design variance $\Varp(\cdot)$. Unfortunately, estimator \eqref{GWest} can be very unstable and take negative values for individual small domains. Therefore, the straightforward estimation of optimal weights \eqref{opt_lambda} is avoided. 

To evaluate the optimal coefficients for compositions \eqref{gen_lin_comb}, one can set a common weight for all domains and then minimize a total MSE with respect to that weight \cite{PK_1979}. A similar approach is to apply James--Stein method \cite[Section 3.4]{RM_2015}. One more idea is SSD estimation \cite{DSC_1982}, where estimators of the weights in \eqref{gen_lin_comb} are taken to be of the form
\begin{equation}\label{ssd_weights}
\hat\lambda_i=\hat\lambda_i(\delta)=
\begin{cases}
1 &\text{if $\widehat N_i/N_i\geq \delta$,}\\
\widehat N_i/(\delta N_i) &\text{otherwise.}
\end{cases}
\end{equation} 
These weights are dependent on the single subjectively chosen parameter $\delta$ for all domains with default value $\delta=1$. Similar SSD estimators were derived in \cite{SH_1989} applying a conditional analysis.

Estimation of MSEs of the design-based composite estimators like these is known as a difficult problem in the literature \cite[Chapter 3]{RM_2015}. One general solution is to treat the composition $\hat\theta_i^{\mathrm C}=\tilde\theta_i^{\mathrm C}(\hat\lambda_i)$ as a synthetic estimator and use the estimator
\begin{equation}\label{GWest_comp}
\mseu (\hat\theta_i^{\mathrm C})=(\hat\theta_i^{\mathrm C}-\hat\theta_i^{\mathrm d})^2-\hat\sigma^2(\hat\theta_i^{\mathrm C}-\hat\theta_i^{\mathrm d})+\hat\sigma^2(\hat\theta_i^{\mathrm C})
\end{equation}
of $\MSEp(\hat\theta_i^{\mathrm C})$, see \cite[Example 3.3.1]{RM_2015} and \cite{C_2020}. However, this estimator has the same drawbacks as \eqref{GWest}. Another general method is to assume that the estimator $\hat\theta_i^{\mathrm C}$ defined by \eqref{gen_lin_comb} approximates the optimal combination $\hat\theta_i^{\mathrm {opt}}=\tilde\theta_i^{\mathrm C}(\lambda_i^*)$ quite well and derive the approximation \cite{C_2021}
\begin{equation*}\label{opt_mse_approx}
\MSEp(\hat\theta_i^{\mathrm C})\approx \lambda_i(1-\lambda_i)\psi_i+\Varp (\hat\theta_i^{\mathrm C})
\end{equation*}
with the empirical version
\begin{equation}\label{mse_est_new}
\mseb (\hat\theta_i^{\mathrm C})=\hat \lambda_i(1-\hat \lambda_i)\hat\psi_i+\hat\sigma^2(\hat\theta_i^{\mathrm C}),
\end{equation} 
where we would set $\hat\psi_i=\hat\psi_i^{\mathrm s}$ to have robust MSE estimators. Estimator \eqref{mse_est_new} takes only non-negative values.

\subsection{Composition based on a ratio of variances}\label{s:42} 

The sampling variance $\psi_i$ is approximately proportional to the product $\theta_i(1-\theta_i)$. That is, one can use the approximation 
\begin{equation}\label{gen_var_approx}
\psi_i\approx \frac{D_i\theta_i(1-\theta_i)}{n_i},
\end{equation}
where $D_i$ is the design effect reflecting the sample efficiency of the complex sampling design \cite{K_1995}. Then, inserting $\hat\theta_i^{\mathrm d}$ and an appropriate estimator $\widehat D_i$ of $D_i$ into \eqref{gen_var_approx}, we would approximate the direct estimator $\hat \psi_i^{\mathrm d}$ of $\psi_i$. Let us first suppose that the domain proportions $\theta_i$ are small, say $\theta_i<0.1$. In that case, it is even more complicated to get reliable direct estimates and estimates of their accuracy \cite{KG_1998}. For example, the direct estimator of the proportion of the unemployed can take zero value even for a sample of moderate size in the municipality. 

Consider two candidate estimators $\hat \psi_i^{\mathrm d}$ and $\hat \psi_i^{\mathrm s}$ of $\psi_i$ used in regression-synthetic estimator \eqref{Rsyn}. Assume that we got too small estimate $\hat\theta_i^{\mathrm d}$ of $\theta_i$ for the specific sample $s$. The direct estimate $\hat \psi_i^{\mathrm d}$ then underestimates the sampling variance $\psi_i$. Therefore, the inequality $\hat \psi_i^{\mathrm s}>\hat \psi_i^{\mathrm d}$ should often hold, that is, the smoothed variance $\hat \psi_i^{\mathrm s}$ could be a better choice than $\hat \psi_i^{\mathrm d}$. Now suppose that $\hat \theta_i^{\mathrm d}$ overestimated the parameter $\theta_i$. Then $\hat \psi_i^{\mathrm d}$ overestimates $\psi_i$ as well, and the inequality $\hat \psi_i^{\mathrm s}<\hat \psi_i^{\mathrm d}$ should hold if $\hat \theta_i^{\mathrm d}$ is an outlier. That larger estimate $\hat \psi_i^{\mathrm d}$ can be employed to down-weight the outlying observation $\hat \theta_i^{\mathrm d}$ used in \eqref{beta_est} thus robustifying synthetic estimators \eqref{Rsyn}. From these considerations, we derive the combined estimators
\begin{equation*}\label{comb}
\hat \psi_i^{\mathrm c}=\max \{ \hat \psi_i^{\mathrm s}, \hat \psi_i^{\mathrm d} \}, \qquad i=1,\ldots, M,
\end{equation*}
of the sampling variances $\psi_i$ that should improve the regression-synthetic estimation. Next, in line with the same ideas, we define the design-based composite estimators
\begin{equation}\label{new_comb}
\hat\theta_i^{\mathrm C}=\hat \lambda_i\hat\theta_i^{\mathrm d}+(1-\hat \lambda_i)\hat\theta_i^{\mathrm S}(\hat \psi_i^{\mathrm c}) \quad \text{with} \quad \hat \lambda_i=\frac{\min \{ \hat \psi_i^{\mathrm s}, \hat \psi_i^{\mathrm d} \}}{\hat \psi_i^{\mathrm c}}, \qquad i=1,\ldots, M,
\end{equation}
of domain proportions \eqref{target_means}. If the estimate $\hat\theta_i^{\mathrm d}$ is an outlier by its small or large value, then relatively more weight is attached to the synthetic part of composition \eqref{new_comb}. The composition is a shrinkage estimator because it shrinks the direct estimator toward the synthetic one.

We apply the same arguments to create \eqref{new_comb} if the parameters $\theta_i$ are not small, but then the inequalities $\max \{\theta_i, \hat\theta_i^{\mathrm d}\}<1/2$ or $\min\{\theta_i, \hat\theta_i^{\mathrm d}\}>1/2$ must be satisfied. If these inequalities are not valid, the composite estimator is still applicable, but it can be less efficient. The worst scenario here would be a large difference $\theta_i-\hat\theta_i^{\mathrm d}$ and the relation $\theta_i\approx 1-\hat\theta_i^{\mathrm d}$ but those events are rare.

To estimate MSE of composition \eqref{new_comb}, we suggest to apply general estimator \eqref{mse_est_new}. We study the accuracy of both these estimators in Section \ref{s:5}.

\subsection{Sample-size-dependent estimation}\label{s:43}

A choice of the parameter $\delta$ in \eqref{ssd_weights} varies from survey to survey. That is, the values 2/3 and 1 are good for LFS in \cite{DSC_1982}, the authors of \cite{UGMS_2009}  try the larger points 1.5 and 2 for their data, and optimal values of $\delta$ are even higher in \cite{C_2020}. Therefore, to select the value of the parameter for the composition $\tilde\theta_i^{\mathrm C}(\delta)=\tilde\theta_i^{\mathrm C}(\hat\lambda_i(\delta))$ defined by \eqref{gen_lin_comb}, we minimize numerically the sample based function \cite{C_2020}
\begin{equation}\label{risk_func}
r(\delta)=\frac{1}{M}\sum_{i=1}^M \mseu (\tilde\theta_i^{\mathrm C}(\delta))
\end{equation}
with respect to $\delta$. This function is the average of individual MSE estimators \eqref{GWest_comp} over domains and therefore it is stable unlike the individual ones. So we get the adaptive composite estimators
\begin{equation}\label{adaptive_comp}
\hat\theta_i^{\mathrm {SSD}}=\tilde\theta_i^{\mathrm C}(\hat\delta^*) \quad \text{with} \quad \hat\delta^*=\argmin_{\delta> 0} r(\delta), \qquad i=1,\ldots, M,
\end{equation}
of the domain proportions. We apply estimators \eqref{mse_est_new} to evaluate MSEs of these compositions.

\section{Simulations using the Labor Force Survey data}\label{s:5}

The main LFS variable is the categorical one that indicates an individual's participation in the labor market. This variable is decomposed into three binary variables: is the person unemployed, is employed, and is not in the labor force. We are going to estimate the proportions of the former two variables in the municipalities of Lithuania. To imitate the real survey, we construct the artificial population from the sample data of the fourth quarter of 2018 as follows: we remove municipalities with too small fractions of observed unemployed persons and then replicate the data of each individual the number of times equal to the rounded survey weight. The size of that population ${\cal U}$ is $N=1396763$, and it contains $M=30$ municipalities. In LFS, the sample of households is drawn without replacement with probabilities proportional to the number of their members, and then the selected households are surveyed entirely. We use the same sampling design to draw $R=10^3$ independent samples of households of size $n'=3700$, and it yields the samples of persons of sizes close to $n=7667$. Then, for the $k$th individual that belongs to the $l$th household of size $h_l$, we apply the approximation $\pi_k \approx h_ln'/N$, $k\in {\cal U}$.

We compare the direct estimator $\hat \theta_{i}^{\mathrm d}=\hat \theta_{i}^{\mathrm H}$ from \eqref{dir_ratio}, regression-synthetic estimator \eqref{Rsyn}, EBLUP \eqref{eblup} calculated using the package \pkg{sae} for \pkg{R} \cite{MM_2015}, and two design-based compositions \eqref{new_comb} and \eqref{adaptive_comp}. Moreover, we compare the accuracy of MSE estimator \eqref{mse_eblup} for \eqref{eblup} with that of two MSE estimators \eqref{GWest_comp} and \eqref{mse_est_new} applied to both compositions \eqref{new_comb} and \eqref{adaptive_comp}. We consider also the optimal combination $\hat\theta_i^{\mathrm {opt}}$ that uses \eqref{opt_lambda} in \eqref{gen_lin_comb}, and its MSE estimator calculated by \eqref{mse_est_new}. 

To model the direct estimates of the proportions of interest by \eqref{M1_eblup} and \eqref{M1}, we use the municipality characteristics $\mathbf{z}_i=(1, z_{i2}, z_{i3}, z_{i4}, z_{i5}, z_{i6})'$, where $z_{i2}$ is the proportion of registered unemployed individuals derived from the administrative Lithuanian Labor Exchange data, $z_{i3}$ is the proportion of persons who, according to the register of the State Social Insurance Fund Board, paid the social contribution one month before they participated in the survey, $z_{i4}$ is the proportion of males, and $z_{i5}$ and $z_{i6}$ are the proportions of individuals from age intervals 26--40 and 41--55, respectively. 

Since the sampling fractions are small in the municipalities, we take $\pi_{kl}\approx\pi_k\pi_l$, $k\neq l$, and so approximate direct estimators \eqref{dir_ratio_var_est} of sampling variances \eqref{dir_ratio_var} by
\begin{equation*}
\hat \psi_i^{\mathrm H}\approx\hat \psi_i^{\mathrm d}=\frac{1}{\widehat N_i^2} \sum_{k\in s_i} w_k(w_k-1) (y_k-\hat \theta_{i}^{\mathrm d})^2, \qquad i=1,\ldots, M,
\end{equation*}
where we write $w_k=1/\pi_k$. Then we smooth these $\hat \psi_i^{\mathrm d}$ to obtain $\hat \psi_i=\hat \psi_i^{\mathrm {sD}}$ according to \eqref{smooth_Dick} and use the smoothed estimates for \eqref{eblup}, \eqref{mse_eblup}, \eqref{Rsyn}, \eqref{mse_est_new}, and in the synthetic parts of \eqref{adaptive_comp} and $\hat\theta_i^{\mathrm {opt}}$.

We apply the bootstrap method of \cite{RWY_1992} to evaluate the estimators of the design variances in \eqref{GWest_comp}, \eqref{mse_est_new}, and \eqref{risk_func}. Denote by $\hat\theta_i$ any estimator for which we need to estimate the design variance. The bootstrap procedure works as follows: (i) Draw a simple random sample of $m=n'-1$ households with replacement from $n'$ sample households. Let $m_l^*$ be the number of times the $l$th sample household is selected, and then $\sum_{l=1}^{n'}m_l^*=m$. Define the bootstrap weights $w_l^*=n'm_l^*w_l/m$, $l=1,\ldots, n'$. Calculate the bootstrap estimate $\hat\theta_i^*$ using the weights $w_l^*$ in the formula for $\hat\theta_i$. (ii) Repeat step (i) $B$ times independently to obtain the estimates $\hat\theta_i^{*(b)}$, $b=1,\ldots, B$. Then
\begin{equation*}
\hat\sigma^2(\hat\theta_i)=\frac{1}{B}\sum_{b=1}^B (\hat\theta_i^{*(b)}-\bar\theta_i^*)^2, \quad \text{where} \quad \bar\theta_i^*=\frac{1}{B}\sum_{b=1}^B \hat\theta_i^{*(b)},
\end{equation*}
is the bootstrap estimator of the design variance $\Varp (\hat\theta_i)$. We take $B=200$.

We evaluate all estimators for each of $R$ samples and calculate approximations to their root mean squared errors (RMSEs) and absolute biases (ABs). That is, we use the accuracy measures
\begin{equation}\label{acc_char}
\RMSE(\hat\mu_i)=\Biggl(\frac{1}{R}\sum_{r=1}^R (\hat\mu_i^{(r)}-\mu_i)^2\Biggr)^{1/2} \quad \text{and} \quad \AB(\hat\mu_i)=\Biggl|\frac{1}{R}\sum_{r=1}^R \hat\mu_i^{(r)}-\mu_i\Biggr|,
\end{equation}
where $\hat\mu_i^{(r)}$ is a realization of the specific estimator $\hat\mu_i$ of the parameter $\mu_i$, based on the $r$th sample. We classify the municipalities by the expected domain sample size into three classes of equal size, and calculate the average of RMSEs as well as ABs over domains of each class. We also present the averages of \eqref{acc_char} over all municipalities as common indicators of accuracy. 

The results for the proportions of the unemployed and the employed are presented in Tables \ref{tab:1} and \ref{tab:2}, respectively. Let us use the superscripts of estimators to discuss the output. In both the tables, any indirect estimator of the proportions improves the direct one in the sense of RMSE, and theoretical composition opt is the best estimator. Among the indirect estimators, regression-synthetic estimator S has much larger design biases than compositions FH, C, and SSD. In Table \ref{tab:1}, the averages of RMSEs over all domains of design-based composite estimators C and SSD are smaller than that of EBLUP FH. It is not valid for estimator C in Table \ref{tab:2} because the proportions of the employed are distributed near the point $1/2$ if to look at the five-number summary $(0.379, 0.585, 0.634, 0.668, 0.766)$ for the true proportions. 

MSE estimation \eqref{mse_est_new} for design-based compositions C and SSD evidently improves estimation \eqref{GWest_comp}, and yields similar or even better results than MSE estimator \eqref{mse_eblup} for FH. The best MSE estimation using \eqref{mse_est_new} is obtained for optimal composition opt. Composite estimators C and SSD only approximate the optimal one and, therefore, their MSE estimators have larger errors. On the other hand, these errors are acceptable if to compare them with the results for FH.

The same experiment but with the twice smaller sample size $n'=1850$ leads to similar conclusions. In this case, design-based composition C improves EBLUP FH more for small proportions.

\section{Conclusions}\label{s:6}

The construction of composite estimator \eqref{new_comb} is based on the monotonicity of the variance of the direct estimator as the function of the proportion. Approximation \eqref{gen_var_approx} is the monotone function in two separate parts of the interval $[0, 1]$. Therefore, the composition loses its efficiency for the proportions close to turning point $1/2$, where the monotonicity changes. 

In general, the sampling variance of any direct estimator of the domain mean is not the monotone function of the target parameter. On the other hand, some GVF models from \cite[p. 274]{W_2007} suggest that this function might be treated as an approximately monotonic one. Therefore, if we can find the GVF model that fits the data well and is the monotonic function, then estimator \eqref{new_comb} could be also applied to the domain means with this fitted model used instead of smoothed sampling variances \eqref{smooth_Dick}.

The simulation study shows that the design-based compositions might be an alternative to the classical EBLUP estimating proportions in small domains. Adaptive composite estimator SSD works well for both unemployment and employment cases, while simpler composition \eqref{new_comb} is efficient for the unemployment fractions that are small proportions. 

Design-based estimators and estimators of MSE under the design-based approach are desirable in practice \cite{MS_2019}. That design MSE estimator \eqref{mse_est_new} works well in our simulations, and its formula is simple compared to that of model MSE estimator \eqref{mse_eblup} for EBLUP.

\bibliographystyle{NAplain}
\bibliography{Ciginas}

\pagebreak

\begin{table}[!h]
\caption{Average RMSEs and ABs of estimators for the unemployed proportions in domain size classes as $n\approx 7667$. The domain is small if its expected sample size $\bar n_i=\Ep (n_i)<116$, is medium for $116\leq\bar n_i<159$, and is large as $\bar n_i\geq 159$.}\label{tab:1}
\centering
\tabcolsep=9pt 
\vspace{1mm} 
\begin{tabular}{lp{10mm}p{10mm}p{10mm}p{10mm}p{10mm}p{10mm}p{10mm}p{10mm}}
\boldhline
 &                \multicolumn{ 4}{c}{Average RMSE ($\times 10^2$)} &                  \multicolumn{ 4}{c}{Average AB ($\times 10^2$)} \\
\cline{2-9}
Estimator          & \multicolumn{ 4}{c}{Domain size class by $\bar n_i$} & \multicolumn{ 4}{c}{Domain size class by $\bar n_i$} \\

           &        any &         small &      medium &        large &         any &         small &      medium &         large \\
\hline

    $\hat \theta_i^{\mathrm d}$     &     2.4793 &     3.8540 &     2.4578 &     1.1259 &     0.0636 &     0.1200 &     0.0485 &     0.0223 \\

     $\hat\theta_i^{\mathrm S}$   &     1.8174 &     2.8950 &     1.5632 &     0.9940 &     1.3461 &     2.3656 &     1.0677 &     0.6050 \\

     $\hat\theta_i^{\mathrm {FH}}$  &     1.7857 &     2.6707 &     1.7156 &     0.9707 &     0.7349 &     1.4738 &     0.5496 &     0.1811 \\

     $\hat\theta_i^{\mathrm C}$  &     1.7511 &     2.6798 &     1.6838 &     0.8897 &     0.7951 &     1.4777 &     0.6130 &     0.2946 \\

     $\hat\theta_i^{\mathrm {SSD}}$    &     1.7529 &     2.7228 &     1.6162 &     0.9196 &     0.8649 &     1.4974 &     0.6928 &     0.4045 \\

     $ \hat\theta_i^{\mathrm {opt}}$   &     1.4712 &     2.3804 &     1.2486 &     0.7846 &     0.7301 &     1.3978 &     0.5206 &     0.2720 \\

     $\mse (\hat\theta_i^{\mathrm {FH}})$  &     0.0223 &     0.0445 &     0.0173 &     0.0051 &     0.0180 &     0.0373 &     0.0128 &     0.0039 \\

$\mseu (\hat\theta_i^{\mathrm C})$   &     0.0708 &     0.1540 &     0.0491 &     0.0094 &     0.0263 &     0.0532 &     0.0215 &     0.0041 \\

$\mseb (\hat\theta_i^{\mathrm C})$   &     0.0173 &     0.0371 &     0.0119 &     0.0030 &     0.0135 &     0.0296 &     0.0087 &     0.0021 \\

$\mseu (\hat\theta_i^{\mathrm {SSD}})$  &     0.0593 &     0.1164 &     0.0494 &     0.0120 &     0.0115 &     0.0290 &     0.0051 &     0.0005 \\

$\mseb (\hat\theta_i^{\mathrm {SSD}})$   &     0.0257 &     0.0497 &     0.0210 &     0.0063 &     0.0172 &     0.0314 &     0.0153 &     0.0050 \\

$\mseb (\hat\theta_i^{\mathrm {opt}})$   &     0.0098 &     0.0206 &     0.0064 &     0.0023 &     0.0050 &     0.0110 &     0.0027 &     0.0012 \\

\boldhline
\end{tabular}  
\end{table}

\begin{table}[!h]
\caption{Average RMSEs and ABs of estimators for the employed proportions in domain size classes as $n\approx 7667$. The domain is small if its expected sample size $\bar n_i=\Ep (n_i)<116$, is medium for $116\leq\bar n_i<159$, and is large as $\bar n_i\geq 159$.}\label{tab:2}
\centering
\tabcolsep=9pt 
\vspace{1mm} 
\begin{tabular}{lp{10mm}p{10mm}p{10mm}p{10mm}p{10mm}p{10mm}p{10mm}p{10mm}}
\boldhline
 &                \multicolumn{ 4}{c}{Average RMSE ($\times 10^2$)} &                  \multicolumn{ 4}{c}{Average AB ($\times 10^2$)} \\
\cline{2-9}
Estimator          & \multicolumn{ 4}{c}{Domain size class by $\bar n_i$} & \multicolumn{ 4}{c}{Domain size class by $\bar n_i$} \\

           &        any &         small &      medium &        large &         any &         small &      medium &         large \\
\hline

    $\hat \theta_i^{\mathrm d}$     &     4.7718 &     6.9201 &     4.7104 &     2.6848 &     0.1516 &     0.2577 &     0.1395 &     0.0575 \\

     $\hat\theta_i^{\mathrm S}$  &     3.4061 &     4.9905 &     3.2215 &     2.0061 &     2.6481 &     4.1247 &     2.5006 &     1.3188 \\

     $\hat\theta_i^{\mathrm {FH}}$  &     3.3054 &     4.6679 &     3.1768 &     2.0716 &     1.7276 &     2.8992 &     1.6535 &     0.6302 \\

     $\hat\theta_i^{\mathrm C}$  &     4.2265 &     6.0532 &     4.1539 &     2.4724 &     0.4024 &     0.6893 &     0.4213 &     0.0967 \\

     $\hat\theta_i^{\mathrm {SSD}}$    &     3.2747 &     4.7502 &     3.1425 &     1.9314 &     1.6996 &     2.6130 &     1.6237 &     0.8622 \\

     $ \hat\theta_i^{\mathrm {opt}}$   &     2.8602 &     4.1800 &     2.6261 &     1.7746 &     1.6026 &     2.5092 &     1.4717 &     0.8269 \\

     $\mse (\hat\theta_i^{\mathrm {FH}})$   &     0.0724 &     0.1281 &     0.0670 &     0.0221 &     0.0561 &     0.1070 &     0.0469 &     0.0143 \\

$\mseu (\hat\theta_i^{\mathrm C})$   &     0.0666 &     0.1371 &     0.0481 &     0.0144 &     0.0297 &     0.0590 &     0.0216 &     0.0086 \\

$\mseb (\hat\theta_i^{\mathrm C})$   &     0.0535 &     0.1093 &     0.0406 &     0.0107 &     0.0102 &     0.0170 &     0.0120 &     0.0016 \\

$\mseu (\hat\theta_i^{\mathrm {SSD}})$  &     0.1992 &     0.3628 &     0.1784 &     0.0563 &     0.0297 &     0.0674 &     0.0201 &     0.0014 \\

$\mseb (\hat\theta_i^{\mathrm {SSD}})$   &     0.0655 &     0.1091 &     0.0697 &     0.0178 &     0.0442 &     0.0715 &     0.0491 &     0.0119 \\

$\mseb (\hat\theta_i^{\mathrm {opt}})$   &     0.0181 &     0.0374 &     0.0110 &     0.0059 &     0.0082 &     0.0155 &     0.0049 &     0.0043 \\

\boldhline
\end{tabular}  
\end{table}

\end{document}